# Structure of interacting aggregates of silica nanoparticles in a polymer matrix: Small-angle scattering and Reverse Monte-Carlo simulations

Julian Oberdisse[1], Peter Hine[2] and Wim Pyckhout-Hintzen[3]

[1]*Laboratoire des Colloïdes, Verres et Nanomatériaux, UMR 5587
CNRS,Université Montpellier II, 34095 Montpellier, France*
[2]*IRC in Polymer Science and Technology, School of Physics and Astronomy
University of Leeds, Leeds, LS2 9JT, United Kingdom*
[3]*Institut für Festkörperforschung, FZ Jülich, 52425 Juelich, Germany*

13[th] of October 2006

**Figures : 10**
**Tables :  3**

**ABSTRACT**

Reinforcement of elastomers by colloidal nanoparticles is an important application where microstructure needs to be understood - and if possible controlled – if one wishes to tune macroscopic mechanical properties. Here the three-dimensional structure of big aggregates of nanometric silica particles embedded in a soft polymeric matrix is determined by Small Angle Neutron Scattering. Experimentally, the crowded environment leading to strong reinforcement induces a strong interaction between aggregates, which generates a prominent interaction peak in the scattering. We propose to analyze the total signal by means of a decomposition in a classical colloidal structure factor describing aggregate interaction and an aggregate form factor determined by a Reverse Monte Carlo technique. The result gives new insights in the shape of aggregates and their complex interaction in elastomers. For comparison, fractal models for aggregate scattering are also discussed.

# I. INTRODUCTION

There is an intimate relationship between microscopic structure and mechanical properties of composite materials [1-5]. Knowledge of both is therefore a prerequisite if one wishes to model this link [6-8]. A precise characterization of the three-dimensional composite structure, however, is usually difficult, as it has often to be reconstructed from two-dimensional images made on surfaces, cuts or thin slices, using electron microscopy techniques or Atomic Force Microscopy [9-11]. Scattering is a powerful tool to access the bulk structure in a non-destructive way [12,13]. X-ray scattering is well suited for many polymer-inorganic composites [14-16], but neutron scattering is preferred here due to the extended q-range (with respect to standard x-ray lab-sources), giving access to length scales between some and several thousand Angstroms. Also, cold neutrons penetrate more easily macroscopically thick samples, and they offer the possibility to extract the conformation of polymer chains inside the composite in future work [17]. Small Angle Neutron Scattering (SANS) is therefore a method of choice to unveil the structure of nanocomposites.

This article deals with the structural analysis by SANS of silica aggregates in a polymeric matrix. Such structures have been investigated by many authors, often with the scope of mechanical reinforcement [18-21], but sometimes also in solution [22-24]. One major drawback of scattering methods is that the structure is obtained in reciprocal space. It is sometimes possible to read off certain key features like fractal dimensions directly from the intensity curves, and extensive modeling can be done, e.g. in the presence of a hierarchy of fractal dimensions, using the famous Beaucage expressions [25]. Also, major progress has been made with inversion to real space data [26]. Nonetheless, complex structures like interacting aggregates of filler particles embedded in an elastomer for reinforcement purposes are still an important challenge. The scope of this article is to report on recent progress in this field.

## II. MATERIALS AND METHODS

### II.1 Sample preparation.

We briefly recall the sample preparation, which is presented in [27]. The starting components are aqueous colloidal suspensions of silica from Akzo Nobel (Bindzil 30/220 and Bindzil 40/130), and nanolatex polymer beads. The latter was kindly provided by Rhodia. It is a core-shell latex of randomly copolymerized Poly(methyl methacrylate) (PMMA) and Poly(butylacrylate) (PBuA), with some hydrophilic polyelectrolyte (methacrylic acid) on the surface. From the analysis of the form factors of silica and nanolatex measured separately by SANS in dilute aqueous solutions we have deduced the radii and polydispersities of a log-normal size distribution of spheres [27]. The silica B30 has an approximate average radius of 78 Å (resp. 96 Å for B40), with about 20% (resp. 28%) polydispersity, and the nanolatex 143 Å (24% polydispersity).

Colloidal stock solutions of silica and nanolatex are brought to desired concentration and pH, mixed, and degassed under primary vacuum in order to avoid bubble formation. Slow evaporation of the solvent at T = 65°C under atmospheric pressure takes about four days, conditions which have been found suitable for the synthesis of smooth and bubble-free films without any further thermal treatment. The typical thickness is between 0.5 and 1 mm, i.e. films are macroscopically thick.

**II.2 Small Angle Neutron Scattering.**

The data discussed here have been obtained in experiments performed at ILL on beamline D11 [27]. The wavelength was fixed to 10.0 Å and the sample-to-detector distances were 1.25 m, 3.50 m, 10.00 m, 36.70 m, with corresponding collimation distances of 5.50 m, 5.50 m, 10.50 m and 40.00 m, respectively. Primary data treatment has been done following standard procedures, with the usual subtraction of empty cell scattering and $H_2O$ as secondary calibration standard [12]. Intensities have been converted to $cm^{-1}$ using a measurement of the direct beam intensity. Background runs of pure dry nanolatex films show only incoherent scattering due to the high concentration of protons, as expected for unstructured random copolymers. The resulting background is flat and very low as compared to the coherent scattering in the presence of silica, and has been subtracted after the primary data treatment.

**III. STRUCTURAL MODELLING**

**III.1 Silica-latex model nanocomposites.**

We have studied silica-latex nanocomposites made by drying a mixture of latex and silica colloidal solutions. The nanometric silica beads can be kept from aggregating during the drying process by increasing the precursor solution pH, and thus their electric charge. Conversely, aggregation can be induced by reducing the solution pH. The resulting nanocomposite has been shown to have very interesting mechanical properties even at low filler volume fraction. The reinforcement factor, e.g., which is expressed as the ratio of Youngs modulus of the composite and the one of its matrix, $E/E_{latex}$, can be varied by a factor of several tens at constant volume fraction of silica (typically from 3 to 15%) [28,29]. In this context it is important to recognize that the silica-polymer interface is practically unchanged from one sample to the other, in the sense that there are no ligands or grafted chains connecting the silica to the matrix. There might be changes to the presence of ions, but their impact on the reinforcement factor appears to be of $2^{nd}$ order [30]. Possible changes in the matrix properties are cancelled in the reinforcement factor representation, the influence of the silica structure is thus clearly highlighted in our experiments. Using a simplified analysis of the structural data measured by SANS, we could show that **(i)** the silica bead aggregation was indeed governed by the solution pH, and **(ii)** the change in aggregation number $N_{agg}$ was accompanied by a considerable change in reinforcement factor at constant silica volume fraction. Although we had convincing evidence for aggregation, it seemed difficult to close the gap and verify that the estimated $N_{agg}$ was indeed compatible with the measured intensity curves. This illustrates one of the key problems in the physical understanding of the reinforcement effect: interesting systems for reinforcement are usually highly crowded, making structural analysis complicated and thereby impeding the emergence of a clear structure-mechanical properties relationship. It is the scope of this article to propose a method for structural analysis in such systems.

**III.2 Modelling the scattered intensity for interacting aggregates.**

For monodisperse silica spheres of volume $V_{si}$, the scattered intensity due to some arbitrary spatial organization can be decomposed in the product of contrast $\Delta\rho$, volume fraction of spheres $\Phi$, structure factor, and the normalized form factor of individual spheres, $P(q)$ [12, 13]. If in addition spheres are organized in monodisperse aggregates, the structure factor can be separated in the intra-aggregate structure factor $S_{intra}(q)$, and a structure factor describing the center-of-mass correlations of aggregates, $S_{inter}(q)$:

$$I(q) = \Delta\rho^2 \ \Phi \ V_{si} \ S_{inter}(q) \ S_{intra}(q) \ P(q) \qquad (1)$$

Here the product $S_{intra}(q) P(q)$ can also be interpreted as the average form factor of aggregates, as it would be measured at infinite dilution of aggregates. In order to be able to compare it to the intensity in $cm^{-1}$, we keep the prefactors and define the aggregate form factor $P_{agg} = \Delta\rho^2 \Phi V_{si} S_{intra}(q) P(q)$.

The above mentioned conditions like monodispersity are not completely met in our experimental system. However, it can be considered sufficiently close to such an ideal situation for this simple scattering law to be applicable. The small polydispersity in silica beads, e.g., is not expected to induce specific aggregate structures. At larger scale, the monodispersity of the aggregates is a working hypothesis. It is plausible because of the strong scattering peak in $I(q)$, which will be discussed with the data. Strong peaks are usually associated with ordered and thus not too polydisperse domain sizes [31].

To understand the difficulty of the structural characterization of the nanocomposites discussed here, one has to see that aggregates of unknown size interact with each other through an unknown potential, which determined their final (frozen) structure. Or from a more technical point of view, we know neither the intra- nor the inter-aggregate structure factor, respectively denoted $S_{intra}(q)$ (or equivalently, $P_{agg}(q)$), and $S_{inter}(q)$.

In the following, we propose a method allowing the separation of the scattered intensity in $P_{agg}(q)$ and $S_{inter}(q)$, on the assumption of **(a)** a (relative) monodispersity in aggregate size, and **(b)** that $P_{agg}$ is smooth in the q-range around the maximum of $S_{inter}$. The inter-aggregate structure factor will be described with a well-known model structure factor developed for simple liquids and applied routinely to repulsively interacting colloids [32-34]. The second factor of the intensity, the aggregate form factor, will be analyzed in two different ways. First, $P_{agg}$ will be compared to fractal models [25]. Then, in a second part, its modeling in direct space by Reverse Monte Carlo will be implemented and discussed [35-39].

**Determination of the average aggregation number and $S_{inter}$.**

Aggregation number and aggregate interaction need to be determined first. The silica-latex nanocomposites discussed here have a relatively well-ordered structure of the filler phase, as can be judged from the prominent correlation peak in I(q), see Fig. 1 as an example for data. The peak is also shown in the upper inset in linear scale. The position of this correlation peak $q_o$ corresponds to a typical length scale of the sample, $2\pi/q_o$, the most probable distance between aggregates. As the volume fraction (e.g., $\Phi = 5\%$ in Fig.1) and the volume of the elementary silica filler particles $V_{si}$ are known, one can estimate the average aggregation number:

$$N_{agg} = (2\pi/q_o)^3 \, \Phi/V_{si} \qquad (2)$$

Two ingredients are necessary for the determination of the inter aggregate structure factor. The first one is the intensity in absolute units, or alternatively the independent measurement of scattering from isolated silica particles, i.e. at high dilution and under known contrast conditions and identical resolution. The second is a model for the structure factor of objects in repulsive interaction. We have chosen a well-known quasi-analytical structure factor based on the Rescaled Mean Spherical Approximation (RMSA) [33,34]. Originally, it was proposed for colloidal particles of volume V, at volume fraction $\Phi$, carrying an electrostatic charge Q, and interacting through a medium characterized by a Debye length $\lambda_D$. In the present study, we use this structure factor as a parametrical expression, with Q and $\lambda_D$ as parameters tuning the repulsive potential. The Debye length, with represents the screening in solutions, corresponds here to the range of the repulsive potential, whereas Q allows to vary the intensity of the interaction. Although the spatial organization of the silica beads in the polymer matrix is due to electrostatic interactions in solution before film formation, we emphasize that this original meaning is lost in the present, parametrical description.

For the calculation of $S_{inter}$, $\Phi$ is given by the silica volume fraction, and the aggregate volume $V = 4\pi/3 \ R_e^3$ by $N_{agg} \ V_{si}$, with $N_{agg}$ determined by eq.(2). $R_e$ denotes the effective radius of a sphere representing an aggregate. In principle, we are thus left with two parameters, Q and $\lambda_D$. The range $\lambda_D$ must be typically of the order of the distance between the surfaces of neighboring aggregates represented by effective charged spheres of radius $R_e$, otherwise the structure factor would not be peaked as experimentally observed. As a starting value, we have chosen to set $\lambda_D$ equal to the average distance between neighboring aggregate surfaces. We will come back to the determination of $\lambda_D$ below, and regard it as fixed for the moment. Then only the effective charge Q remains to be determined.

Here the absolute units of the intensity come into play. $N_{agg}$ is known from the peak position, and thus also the low-q limit of $S_{intra}(q \to 0)$, because forward scattering of isolated objects gives directly the mass of an aggregate [12]. The numerical value of the (hypothetical) forward scattering in the absence of interaction can be directly calculated using eq.(1), setting $S_{intra} = N_{agg}$ and $S_{inter} = 1$. Of course the aggregates in our nanocomposites are not isolated, as their repulsion leads to the intensity peak and a depression of the intensity at small angles. The limit of $I(q \to 0)$ contains thus also an additional factor, $S_{inter}(q \to 0)$. In colloid science, this factor is known as the isothermal osmotic compressibility [12], and here its equivalent can be deduced from the ratio of the isolated aggregate limit of the intensity ($S_{intra} = N_{agg}$, $S_{inter}$

= 1), and the experimentally measured one $I(q \rightarrow 0)$. It characterizes the strength of the aggregate-aggregate interaction.

Based on the RMSA-structure factor [33,34], we have implemented a search routine which finds the effective charge Q reproducing $S_{inter}(q \rightarrow 0)$. With $\lambda_D$ fixed, we are left with one free parameter, Q, which entirely determines the q-dependence of the inter-aggregate structure factor. An immediate cross-check is that the resulting $S_{inter}(q)$ is peaked in the same q-region as the experimental intensity. In Fig. 1, the decomposition of the intensity in $S_{inter}(q)$ and $S_{intra}(q)$ is shown. It has been achieved with an aggregation number of 93, approximately forty charges per aggregate, and a Debye length of 741 Å, i.e. 85% of the average surface-to-surface distance between aggregates, and we come now back to the determination of $\lambda_D$.

In Fig. 2, a series of inter-aggregate structure factors is shown with different Debye lengths: 50%, 85% and 125% of the distance between neighboring aggregate surfaces (872 Å). The charges needed to obtain the measured compressibility are 27, 40 and 64.5, respectively. In Fig. 2, the inter-aggregate structure factors are seen to be peaked in the vicinity of the experimentally observed peak, with higher peak heights for the lower Debye lengths. Dividing the measured intensity $I(q)$ by $\Delta \rho^2 \, \Phi \, V_{si} \, P(q) \, S_{inter}$ yields $S_{intra}$, also presented in the plot. At low-q, these structure factors decrease strongly, then pass through a minimum and a maximum at intermediate q , and tend towards one at large q (not shown). The high-q maximum is of course due to the interaction between primary particles.

In the low-q decrease, it can be observed that a too strong peak in $S_{inter}$ leads to a depression of $S_{intra}$ at the same q-value. Conversely, a peak that is too weak leads to a shoulder in $S_{intra}$. Only at intermediate values of the Debye length (85%), $S_{intra}$ is relatively smooth. In the following, it is supposed that there is no reason for $S_{intra}$ to present artefacts in the decrease from the Guinier regime to the global minimum (bumps or shoulders), and *set* the Debye length to the intermediate value (85%) for this sample. We have also checked that small variations around this intermediate Debye length (80 to 90%) yield essentially identical structure factors, with peak height differences of a view percent. This procedure of adjusting $\lambda_D$ to the value with a smooth $S_{intra}$ has been applied to all data discussed in this paper.

**Fitting $S_{intra}$ using geometrical and fractal models.**

Up to now, we have determined the inter-aggregate structure factor, and then deduced the experimental intra-aggregate structure factor $S_{intra}$ as shown in Fig.2 by dividing the intensity by $S_{inter}$ according to eq.(1). To extract direct-space information from $S_{intra}$ for aggregates of unknown shape, two types of solutions can be sought. First, one can make use of the knowledge of the average aggregation number, and construct average aggregates in real space. This supposes some idea of possible structures, which can then be Fourier-transformed and compared to the experimental result $S_{intra}(q)$. For example, one may try small crystallites [40], or, in another context, amorphous aggregates [41]. Another prominent case is the one of fractal structures, which are often encountered in colloidal aggregation [42 - 44].

Let us quickly discuss the scattering function of finite-sized fractals using the unified law with both Guinier regime and power law dependence [25, 45]. An isolated finite-sized object with fractal geometry described by a fractal dimension d has three distinct scattering domains. At low q (roughly $q < 1/R_g$), the Guinier law reflects the finite size and allows the measurement of the aggregate mass from the intensity plateau, and of the radius of gyration $R_g$ from the low-q decay. At intermediate q $(q > 1/R_g)$, the intensity follows a power law $q^{-d}$ up to the high-q regime $(q > 1/R)$, which contains the shape information of the primary particles (of radius R) making up the aggregate. Generalizations to higher level structures have also been used [46-49]. Here we use a two-level description following Beaucage [25]:

$$I(q)=G_1\cdot\exp\left(-\frac{q^2R_g^2}{3}\right)+B_1\cdot\left\{\frac{\left[\mathrm{erf}\left(qR_g/6^{1/2}\right)\right]^3}{q}\right\}^d\cdot\exp\left(-\frac{q^2R^2}{3}\right)$$

$$+\ G_2\cdot\exp\left(-\frac{q^2R^2}{3}\right)+B_2\cdot\left\{\frac{\left[\mathrm{erf}\left(qR/6^{1/2}\right)\right]^3}{q}\right\}^p \tag{3}$$

Note that there is no interaction term like $S_{inter}$ in eq.(1), and that eq.(3) accounts only for intra-aggregate structure in this case. The first term on the right-hand-side of eq.(3) is the Guinier expression of the total aggregate. The second term, i.e. the first power law, corresponds to the fractal structure of the aggregate, the error function allowing for a smooth cross-over. This fractal law is weighted by the Guinier expression of the second level, which is the scattering of the primary silica particle in our case; this effectively suppresses the fractal law of the first level at high q. This is followed by an equivalent expression of the higher level, i.e. a Guinier law of primary particles followed by the power-law, which is the Porod law of the primary particles in this case.

**Fitting $S_{intra}$ using Reverse Monte Carlo.**

The second solution to extract real-space information from $S_{intra}$ is to fit the intra-aggregate structure factor by a Monte-Carlo approach which we describe here. It has been called Reverse Monte Carlo (RMC) [35-39] because it is based on a feed-back between the structure in direct and reciprocal space, which makes it basically an automatic fitting procedure once the model is defined. The application of RMC to the determination of the aggregate structure from the scattered intensity is illustrated (in 2D) in Fig. 3. RMC was performed with a specially developed Fortran program as outlined in the Appendix. The method consists in generating representative aggregate shapes by moving elements of the aggregate in a random way - these are the Monte Carlo steps -, and calculate the corresponding structure factor at each step. The intensity is then compared to the experimentally measured one, which gives a criterion whether the Monte Carlo step is to be accepted or not. Monte-Carlo steps are repeated until no further improvement is obtained. If the algorithm converges, the outcome is a structure compatible with the scattered intensity. As an immediate result, it allows us to verify that an aggregate containing $N_{agg}$ filler particles - $N_{agg}$ being determined from the peak position $q_o$ - produces indeed the observed scattered intensity.

## IV. APPLICATION TO EXPERIMENTAL RESULTS

### IV.1 Moderate volume fraction of silica ($\Phi = 5\%$, B30).

**Aggregate interaction.**

We now apply our analysis to the measured silica-latex nanocomposite structures [27]. We start with the example already discussed before (Figs. 1 and 2), i.e. a sample with a moderate silica volume fraction of 5%, and neutral solution pH before solvent evaporation. From the peak position ($q = 3.9 \ 10^{-3} \ \text{Å}^{-1}$), an average aggregation number of $N_{agg} = 93$ can be deduced using eq.(2). The aggregate mass gives us the hypothetical low-q limit of the intensity for non-interaction aggregates using eq. (1), with $S_{inter} = 1$, of 9550 $\text{cm}^{-1}$. The measured value being much lower, approximately 450 $\text{cm}^{-1}$, with some error induced by the extrapolation, the isothermal compressibility due to the interaction between aggregates amounts to about 0.05. This rather low number expresses the strong repulsive interaction. The charged spheres representing the aggregates in the inter-aggregate structure factor calculation have the same volume as the aggregates, and thus an equivalent radius of $R_e = 367$ Å. The surface-to-surface distance between spheres is therefore 872 Å. Following the discussion of Fig. 2, we have set the screening length $\lambda_D$ to 85% of this value, 741 Å. Using this input in the RMSA-calculation, together with the constraint on the compressibility, an electric charge of 40 elementary charges per aggregate is found. The corresponding s $S_{inter}$ are plotted in Fig. 2.

**Fractal modeling.**

A fit with a two level fractal, eq.(3), has been performed with the aggregate form factor $P_{agg}$ obtained by dividing the experimental intensity by $S_{inter}$. The result is shown in Fig. 4. There are several parameters to the fit, some of which can be found independently. The slope of the high-q power law, e.g., has been fixed to p= –4, in agreement with the Porod law. The radius of gyration of the primary particles is 76 Å, and the corresponding prefactor $G_2$ can be deduced from the particle properties [27] and concentration (103 $cm^{-1}$). For comparison, the form factor of the individual particle is shown in Fig. 4 as a one level Beaucage function, i.e. using only the last two terms of eq. (3). Furthermore, we have introduced the $G_1$ value of 9550 $cm^{-1}$ calculated from $N_{agg}$, i.e. from the peak position. Fitting yields the radius of gyration of aggregates (1650 Å), and a fractal dimension of 1.96. At intermediate q, however, the quality of the fit is less satisfying. The discrepancy is due to the minimum of $S_{intra}$ (cf. Fig. 2) around 0.02 $\text{Å}^{-1}$, a feature which is not captured by the model used here (eq. (3)).

**Reverse Monte Carlo.**

We now report on the results of the implementation of an RMC-routine applied to the structure of the sample discussed above ($\Phi = 5\%$, pH 7). In Fig. 5, we plot the evolution of $\chi^2$ (cf. appendix) as a function of the number of Monte-Carlo tries for each bead (on average), starting from the a random initial condition as defined in the appendix. For illustration purposes, this is compared to the $\chi^2$ from different initial conditions, i.e. aggregates constructed according to the same rule but with a different random seed. Such initial aggregate structures are also shown on the left-hand side of Fig. 6. In all cases, the $\chi^2$ value is seen to decrease in Fig. 5 by about two orders of magnitude within five Monte-Carlo steps per bead. It then levels off to a plateau, around which it fluctuates due to the Boltzmann criterion. We have checked that much longer runs do not further increase the quality of the fit, cf. the inset of Fig. 5. The corresponding aggregates at the beginning and at the end of the simulation run are also shown in Fig.6. They are of course different depending on the initial condition and angle of view, but their statistical properties are identical, otherwise their Fourier transform would not fit the experimental data. It is interesting to see how much the final aggregate structures, rather elongated, look similar.

Having established that the algorithm robustly produces aggregates with similar statistical properties, we now compare the result to the experimental intensity in Fig. 7. Although some minor deviations between the intensities are still present, the agreement over five decades in intensity is quite remarkable. It shows that the aggregation number determined from the peak position $q_o$ is indeed a reasonable value, as it allows the construction of a representative aggregate with almost identical scattering behavior. In the lower inset of Fig.7, the RMC result for the aggregate form factor $P_{agg}$ is compared to the experimental one (obtained by dividing the $I(q)$ of Fig.7 by $S_{inter}$). The fit is good, especially as the behavior around 0.02 $\text{Å}^{-1}$ is better described than in the case of the fractal model, Fig. 4.

The radius of gyration can be calculated from the position of the primary particles in one given realization. We find $R_g$ around 1150 Å, a bit smaller than with the fractal model (1650 Å), a difference probably due to the fact that we are only approaching the low-q plateau. For the comparison of the fractal model to RMC, let us recall that both apply only to $P_{agg}$, i.e. after the separation of the intensity in aggregate form factor $P_{agg}$ and structure factor $S_{inter}$. Both methods give the same fractal dimension d of aggregates because this corresponds to the same slope of $P_{agg}$. The aggregate form factor $P_{agg}$ and thus the intensity are better (although not perfectly) fitted with RMC. This is true namely for the minimum around 0.02 Å$^{-1}$, presumably because the nearest neighbor correlations inside each aggregate are captured by a physical model of touching beads. Last but not least, RMC gives snapshots of 3D real-space structures compatible with the scattered intensity, which validates the determination of $N_{agg}$ using eq. (2).

For the sake of completeness, we have tested RMC with aggregation numbers different from the one deduced from the peak position. Taking a very low aggregation number (i.e., smaller than the value obtained with eq.(2))) leads to bad fits, whereas higher aggregation numbers give at first sight acceptable fits. The problem with too high aggregation numbers is that the peak position of $S_{inter}$ is different from the position of the intensity peak due to conservation of silica volume. RMC compensates for this by introducing an oscillation in $S_{intra}$ (or equivalently, $P_{agg}$) which effectively shifts the peak to its experimentally measured position. In the upper inset of Fig.7 $P_{agg}$ presenting such an artefact ($N_{agg}$ = 120 and 150) is compared to the one with the nominal aggregation number, $N_{agg}$ = 93 (filled symbols). The oscillation around 0.004 Å$^{-1}$ is not present with $N_{agg}$ = 93, and becomes stronger as the aggregation number deviates more from the value determined from the intensity peak position, eq.(2).

**IV.2 Evolution with silica volume fraction.**

In the preceding section we have analyzed a sample at moderate silica volume fraction, 5%. It is now interesting to check if the same type of modeling can be applied to higher silica volume fractions and bigger aggregates (i.e., lower solution pH), where the structure factor can be seen to be more prominent directly from I(q).

**Evolution of structure with silica volume fraction ($\Phi$ = 5 and 10%, B30).**

In Fig. 8, two data sets corresponding to a lower pH of 5, for $\Phi$ = 5% and 10% (symbols) are compared to their RMC fits, in linear representation in order to emphasize the peaks. The parameters used for these calculations are given in Table 1, together with the aggregation numbers deduced from the peak position (using eq. (2)). As expected, these are considerably higher than at pH 7 [27]. Concerning the Debye length, it is interesting to note that its value relative to the inter-aggregate distance increases with volume fraction. As we have seen in section III.2, a higher Debye length leads to a weaker peak. This tendency is opposite to the influence of the volume fraction, and we have checked that the peak in $S_{inter}$ is comparable in height in both cases, i.e. the two tendencies compensate.

At first sight of Fig. 8, it is surprising that the intensity at 10% is lower than the one at 5%. This is only true at small-q – the 10% intensity being higher in the Porod domain, as it should, cf. $P_{agg}$ shown in the inset in log-scale. At both concentrations, the aggregate shape seems to be unchanged, (similar fractal dimension d, 2.25 and 2.3 for 5% and 10%, respectively), and together with the shift in peak position by a factor $2^{\frac{1}{3}}$ (as $\Phi$ is doubled) to a region where $P_{agg}$ is much lower, it explains the observed decrease in intensity. We will see in the discussion of a series with the silica B40 that this behavior is not general, and that aggregation depends (as observed before [27]) on the type of bead.

For illustration, the scattered intensity corresponding to the random initial condition of RMC (cf. appendix) is also shown in Fig. 8. The major initial deviation from the experimental values underlines the capacity of the RMC algorithm to converge quickly (cf. Fig. 5) towards a very satisfying fit of the experimental intensity. Note that there is a small angle upturn for the sample at 10%. This may be due to aggregation on a very large scale, which is outside the scope and the possibilities of our method.

**Evolution of structure with silica volume fraction (Φ = 3% - 15%, B40)**

We now turn to a series of samples with a different, slightly bigger silica beads (denoted B40), in a highly aggregated state (low pH), with a larger range of volume fractions. In Fig. 9 the intensities are plotted with the RMC fits, for the series Φ = 3 – 15%, at pH 5, silica B40. The parameters used for the calculations are given in the Table 2.

The fits shown in Fig. 9 are very good, which demonstrates that the model works well over a large range of volume fractions, i.e. varying aggregate-aggregate interaction. Concerning the parameters Debye length and charge, we have checked that the peaks in $S_{inter}$ are comparable in height (within 10%). Only their position shifts, as it was observed with the smaller silica (B30). Unlike the case of B30, however, the intensities follow a 'normal' increase with increase in volume fraction, which suggests a different evolution in aggregate shape and size for the bigger beads.

The case of the lowest volume fraction, Φ = 3%, deserves some discussion. The aggregation number is estimated to 188 using eq. (2). The peak is rather weak due to the low concentration, and it is also close to the minimum q-value. We thus had to base our analysis on an estimation of I(q→0), 700 cm$^{-1}$. The resulting inter-aggregate structure factor $S_{inter}$ is as expected only slightly peaked (peak height 1.1). We found that some variation of $N_{agg}$ does not deteriorate the quality of the fit, i.e. small variations do not introduce artificial oscillations in the aggregate form factor. We have, e.g., checked that the aggregate form factors $P_{agg}$ for $N_{agg} = 120$ and 200 are equally smooth. At higher/lower aggregation number, like 100 or 230, oscillations appear in $P_{agg}$. It is concluded that in this rather dilute case the weak ordering does not allow for a precise determination of $N_{agg}$. For higher volume fractions, $\Phi > 3\%$, the aggregation numbers given in Table 2 are trustworthy.

## V. DISCUSSION

### V.1 Uniqueness of the solution.

The question of the uniqueness of the solution found by RMC arises naturally. Here two different levels need to be discussed. The first one concerns the separation in aggregate form and structure factor. We have shown that the aggregate parameters ($N_{agg}$, aggregate interaction) are fixed by the boundary conditions. Only in the case of weak interaction ($\Phi = 3\%$), acceptable solutions with quite different aggregation numbers (between about 120 and 200) can be found. In the other cases, variations by some 15% in $N_{agg}$ lead to bad intensity fits or artefacts in the aggregate form factor $P_{agg}$. We can thus confirm that one of the main objectives is reached, namely that it is possible to find an aggregate of well-defined mass (given by eq.(2)), the scattering of which is compatible with the intensity.

The second level is to know to what extend the RMC-realizations of aggregates are unique solutions. It is clear from the procedure that many similar realizations are created as the number of Monte Carlo steps increases (e.g., the plateau in Fig. 5), all with a comparable quality of fit. In Fig. 5, this is also seen to be independent from the initial condition, and Figs. 6 and 8 illustrated how far this initial condition is from the final structure. All the final realizations have equivalent statistical properties, and they can be looked at as representatives of a class of aggregates with identical scattering. However, no unique solution exists.

## V.2 From aggregate structure to elastomer reinforcement.

We have shown in previous work that the mechanical properties of our nanocomposites depend strongly on aggregation number and silica volume fraction [28-30]. The aggregation number was estimated from the peak position, and we have now confirmed that such aggregates are indeed compatible with the complete scattering curves. It is therefore interesting to see how the real-space structures found by our method compare to the mechanical properties of the nanocomposites.

The low deformation reinforcement factors of the series in silica volume fraction (B40, pH5, $\Phi = 3 - 15\%$) are recalled in Table 3 [30]. $E/E_{latex}$ is found to increase considerably with $\Phi$, much more than $N_{agg}$. Aggregate structures as resulting from the RMC-procedure applied to the data in Fig. 9 are shown in Fig. 10. At low $\Phi$, aggregates are rather elongated, and with increasing $\Phi$, they are seen to become slightly bulkier. We have determined their radii of gyration and fractal dimension with a one-level Beaucage fit, using only the first two terms of the right-hand-side of eq. (3), and applying the same method as in section IV.1. The results are summarized in Table 3. The fractal dimension is found to increase with $\Phi$, as expected from Fig. 10. The aggregate radius $R_g$ first decreases, then increases again. If we compare $R_g$ to the average distance between aggregates D (from the peak position of $S_{inter}$), we find a crowded environment. The aggregates appear to be tenuous structures, with an overall radius of gyration *bigger than* the average distance between aggregates, which suggests aggregate interpenetration.

In a recent article [30], we have determined the effective aggregate radius and fractal dimension from a mechanical model relating $E/E_{latex}$ to the compacity of aggregates. The numerical values are different (aggregate radii between 1200 and 980 Å, fractal dimensions between 2.1 and 2.45) due to the mechanical model which represents aggregates as spheres, but the tendency is the same: Radii decrease as $\Phi$ increases, implying bulkier aggregates with higher fractal dimensions. Only the increase in radius found at 15% is not captured by the mechanical model.

Our picture of reinforcement in this system is the based on the idea of percolation of hard silica structures in the matrix. Due to the (quasi-)incompressibility of the elastomer matrix, strain in any direction is accompanied by lateral compression, thus pushing aggregates together and creating mechanical percolation. Aggregates are tenuous, interpenetrating structures. The higher the silica volume fraction, the more compact the aggregates (higher d), and the stronger the percolating links. At low $\Phi$, $N_{agg}$ is more or less constant, which implies that the aggregates decrease in size, cf. Table 3 for both fractal and RMC-analysis. Above 6%, $N_{agg}$ increases, and the aggregates become both denser and grow again in size. At the same time, aggregates come closer (D goes down). This moves the system closer to percolation, and leads to the important increase in the reinforcement factor. In other systems, this is also what the reinforcement curves as a function of filler volume fraction suggest [28], where extremely strong structures made of the percolating hard filler phase are found above a critical volume fraction [50].

## VI. CONCLUSION

We have presented a complete analysis of the scattering function of complex spectra arising from strongly aggregated and interacting colloidal silica aggregates in nanocomposites. The main result is the validation of the determination of the average aggregation number by a complete fit of the data. This is achieved by a separation of the scattered intensity in a product of aggregate form and structure factor. The aggregate form factor can then be described either by a fractal model, or by Reverse Monte Carlo modeling. The use of the decomposition of I(q) in a product is based on the assumption that aggregates are similar in size. This is justified by the strong peak in intensity, which indicates strong ordering, incompatible with too high polydispersity in size.

Fractal and RMC-modelling appear to be complementary, with the advantage of generality and simplicity for the fractal model, whereas RMC needs numerical simulations adapted to each case. However, RMC does not rely on approximations (Guinier), and by its geometrical construction it connects local configurations (bead-bead) to the global structure. RMC thus gives a real space picture of aggregates compatible with I(q), and thereby confirms calculation of aggregation numbers from the peak positions.

To finish, possible improvements of our method can be discussed. Technically, the introduction of the spectrometer resolution function is straightforward but would not fundamentally change results, and considerably slow down the algorithm. A more ambitious project is be to get rid of the separation in aggregate form and structure factor by performing a RMC-simulation of a large system containing many aggregates [51]. It will be interesting to see if the Monte-Carlo algorithm converges spontaneously towards more or less monodisperse aggregates, or if very different solutions, not considered in the present work, exist.

**Acknowledgements :** Work conducted within the scientific program of the European Network of Excellence *Softcomp*: 'Soft Matter Composites: an approach to nanoscale functional materials', supported by the European Commission. Silica and latex stock solutions were a gift from Akzo Nobel and Rhodia. Help by Bruno Demé (ILL, Grenoble) as local contact on D11 and beam time by ILL is gratefully acknowledged, as well as support by the instrument responsible Peter Lindner. Thanks also to Rudolf Klein (Konstanz) for fruitful discussions on structure factors.

**APPENDIX:** Reverse Monte Carlo algorithm for scattering from aggregates.

## A.1 Initial aggregate construction

The first step is to build an initial aggregate which can then evolve according to the Monte-Carlo rules in order to fit the experimental intensity I(q) of nanocomposites. From the intensity peak position and eq.(2), the aggregation number $N_{agg}$ is known. The primary particles are the silica beads with a radius drawn from a size distribution function [27]. The initial aggregate is constructed by adding particles to a seed particle placed at the origin. Each new particle is positioned by randomly choosing one of the particles which are already part of the aggregate, and sticking it to it in a random direction. Then, collisions with all particles in the aggregate at this stage are checked, and the particle is accepted if there are no collisions. This is repeated until $N_{agg}$ is reached. Two realizations of initial aggregate structures are shown in Fig. 6.

## A.2 Monte-Carlo steps

The Monte-Carlo steps are designed to change the shape of the aggregate, in order to reach closer agreement with the scattering data. To do this, the local aggregate topology has to be determined. The aim is to identify particles which can be removed from the aggregate without breaking it up, i.e. particles which sit on the (topological) surface of the aggregate. Moving such particles to another position in the aggregate leads to a new structure with updated topology. A Monte-Carlo step thus consists in randomly choosing one of the particles which can be removed, and repositioning it in contact with some other, randomly chosen particle, again in a random direction. As before, it is checked that there are no collisions with the other particles of the aggregate.

**A.3 Fit to experimental intensity**

Each Monte-Carlo step is evaluated by the calculation of the orientationally averaged aggregate form factor $P_{agg}(q)$ , which is multiplied by $S_{inter}(q)$, cf. eq. (1), and compared to the experimental intensity I(q). The comparison is done in terms of $\chi^2$:

$$\chi^2 = \frac{1}{N} \sum_i \left( \frac{I(q_i) - I_{RMC}(q_i)}{\sigma} \right)^2 \qquad (A.1)$$

where the difference between RMC-prediction and experimental intensity is summed over the N q-values. The statistical error $\sigma$ was kept fixed in all calculations. In the our algorithm, the move is accepted if it improves the agreement between the theoretical and experimental curves, or if the increase in $\chi^2$ is moderate in order to allow for some fluctuations. This is implemented by a Boltzmann criterion on $\chi^2$:

$$\exp(-\Delta\chi^2 / B) > \text{random number in the interval } [0,1] \qquad (A.2)$$

In the present implementation, B has been fixed to at most 1% of the plateau value of $\chi^2$. This plateau-value was found to be essentially independent of the choice of B. Given the quality of the fits, a simulated annealing approach was therefore not necessary.

**References:**

[1] *Science and Technology of Rubber*, J.E .Mark, B. Erman, F.R. Eirich, eds, Academic Press, San Diego, 1994

[2] *Mechanical Properties of Polymers and Composite,* L.E. Nielsen, R.F. Landel, Marcel Dekker, New York, 1994

[3] J. Frohlich, W. Niedermeier W, H.D. Luginsland, *Composites A*, 2005, **36** (4), 449

[4] G.R. Pan, J.E. Mark, D.W. Schaefer, *J. Polym. Sci. B*, 2003, **41** (24), 3314

[5] R. Inoubli, S. Dagréou, A. Lapp, L. Billon, J. Peyrelasse, *Langmuir*, 2006, **22**, 6683

[6] T.A. Witten, M. Rubinstein, R.H.J. Colby, *J Phys II (France),* 1993, **3**, 367

[7] G. Heinrich, M. Klüppel, Th. A. Vilgis, *Current Op. Solid State Mat Sci*, 2002, **6**, 195

[8] G. Huber, Th. A. Vilgis, *Macromolecules*, 2002, **35**, 9204

[9] A.I. Medalia, *Rub. Chem. Tech.,* 1974, **47**, 411

[10] H. Bodiguel, H. Montes, C. Fretigny, *Rev. Sci. Instr.*, 2004, **75** (8), 2529

[11] Y. Le Diagon, PhD thesis, University of Paris VI , 2005

[12] *Neutrons, X-ray and Light: Scattering Methods Applied to Soft Condensed Matter*; P. Lindner, Th. Zemb; eds.; North Holland, 2002.

[13] H. Peterlik, P. Fratzl, *Monatshefte für Chemie*, 2006, **137**, 529

[14] D.W.Schaefer, C. Suryawanshi, P. Pakdel, J. Ilavsky, P.R. Jemian, *Physica A*, 2002, **314**, 686

[15] T. Koga,T. M. Takenaka, K. Aizawa,M. Nakamura, T. Hashimoto, *Langmuir*, 2005, **21**(24), 11409

[16] T.P. Rieker, M. Hindermann-Bischoff, F. Ehrburger-Dolle, *Langmuir,* 2000, **16**, 5588

[17] A. Botti, W. Pyckhout-Hintzen, D. Richter, V. Urban, E. Straube, *J. Chem. Phys.*, 2006, **124** (17), 174908

[18] J. Persello, J.P. Boisvert JP, A. Guyard, B. Cabane, *J. Phys. Chem. B*, 2004, **108** (28), 9678

[19] J. Berriot, H. Montes, F. Martin, M. Mauger, W. Pyckhout-Hintzen, G. Meier, H. Frielinghaus *Polymer*, 2003, **44**(17), 4909

[20] Y. Chevalier, M. Hidalgo, J.Y. Cavaille, B. Cabane, *Macromolecules,* 1999, **32**(23), 7887

[21] Y. Rharbi, B. Cabane, A. Vacher, M. Joannicot, F. Boué, *Europhys. Lett.,* 1999, **46** (4), 472

[22] V.L. Alexeev, P. Ilekti, J. Persello, J. Lambard, T. Gulik, B. Cabane, *Langmuir*, 1996, **12** (10), 2392

[23] V.L. Alexeev, *J. Coll Int. Sci*, 1998, **206**, 416

[24] K. Wong, P. Lixon, F. Lafuma, P. Lindner, O. Aguerre Charriol, B. Cabane, *J. Coll. Int. Sci.*, 1992, **153**(1), 55

[25] G. Beaucage, *J. Appl. Cryst.*, 1995, **28** , 717

[26] (a) O. Glatter, *J. Appl. Cryst.*, 1977, **10**, 415 (b) O. Glatter, G. Fritz , H. Lindner, J. Brunner-Popela, R. Mittelbach R., R. Strey., U. Stefan, S. U. Egelhaaf, *Langmuir*, 2000, **16**, 8692

[27] J. Oberdisse, B. Demé, *Macromolecules*, 2002, **35**(4), 4397

[28] J. Oberdisse, *Macromolecules*, 2002, **35**, 9441

[29] J. Oberdisse, *Soft Matter*, 2006, **2**, 29

[30] J. Oberdisse, A. El Harrak, G. Carrot, J. Jestin, F. Boué, *Polymer*, 2005, **46**, 6695

[31] B. d'Aguanno, R. Klein, *J. Chem. Soc. Faraday Trans.*, 1991, **87**, 379

[32] J.P.Hansen, I.R. McDonald, *Theory of Simple Liquids*, Academic Press, 1986

[33] J.B.Hayter, J. Penfold, *Mol. Phys.*, 1981, **42**, 109

[34] J.P. Hansen, J.B. Hayter, *Mol. Phys.,* 1982, **46**, 651

[35] R.L. McGreevy, *J Phys: Cond Mat*, 2001, **13**, R877

[36] R.L. McGreevy, P. Zetterström, *Curr Op Solid State Mat Sci*, 2003, **7**, 41

[37] L. Pusztai, H. Dominguez, O.A. Pizio, *J Coll Int Sci*, 2004, **277**, 327

[38] S. Gruner, O. Akinlade, W Hoyer, *J. Phys:Cond. Mat.*, 2006, **18**(20), 4773

[39] L. Pusztai, R.L. McGreevy, *J. Chem. Phys.* 2006, **125**(4), 044508

[40] J. Oberdisse, Y. Rharbi, F. Boué, *Comput. Theor. Polym. Sci.,* 2000, **10**, 207

[41] J.F. Berret, P. Hervé, O. Aguerre-Chariol, J. Oberdisse, *J Phys Chem B,* 2003, **107**(32), 8111

[42] J.L. Burns, Y. Yan, G.J. Jameson, S. Biggs, *Langmuir,* 1997, **13**, 6413

[43] G. Bushell, R. Amal, *J. Coll. Int. Sci*, 2000, **221**, 186

[44] T.P. Rieker, S. Misono, F. Ehrburger-Dolle, *Langmuir*, 1999, **15**, 914

[45] J. Teixeira, *J. Appl. Cryst.*, 1988, **21**, 781

[46] H.K. Kammler, G. Beaucage, R. Mueller, S.E. Pratsinis, *Langmuir*, 2004, **20**, 1915

[47] D.W. Schaefer, T. Rieker, M. Agamalian, J.S. Lin, D. Fischer, S. Sukumaran, C. Chen, G. Beaucage, C. Herd, J. Ivie, *J. Appl. Cryst.*, 2000, **33**, 587

[48] W.E. Smith, C.F. Zukoski, *Langmuir*, 2004, **20**, 11191

[49] G. Belina, V. Urban, E. Straube, W. Pyckhout-Hintzen, M. Klüppel, G. Heinrich, *Macromol. Symp.*, 2003, **200**, 121

[50] E. Chabert, M. Bornert, E. Bourgeat-Lami, J.-Y.Cavaillé, R. Dendievel, C. Gauthier, J.L. Putaux, A. Zaoui, *Mat Sci Eng A,* 2004*,* **381**, 320

[51] This idea was suggested by Prof. L. Pusztai.

**Figure captions**

Figure 1 :     Structure of silica-latex nanocomposite ($\Phi$ = 5%, pH 7, B30) as seen by SANS. The experimental intensity (○) is represented in log scale, and in linear scale in the upper inset. In the lower inset, the two structure factors $S_{inter}$ and $S_{intra}$, are shown. Such a decomposition is the result of our data analysis as described in the text.

Figure 2 :     Structure factors (for $\Phi$ = 5%, pH 7, B30) obtained with different Debye lengths and charges, but identical compressibility. $\lambda_D$ = 436 Å (50%), Q = 64.5 (◇), $\lambda_D$ = 741 Å (85%), Q = 40 (◆). $\lambda_D$ = 1090 Å (125%), Q = 27 (□). In parentheses the Debye lengths as a fraction of the inter-aggregate surface distance (872 Å). In the inset, a zoom on the artefact in $S_{intra}$ observed at 50%, but not at 85%, is shown.

Figure 3 :     Schematic drawing illustrating the Reverse Monte Carlo algorithm applied to the generation of aggregates. An internal filler particle like the black bead can not be removed without destroying the aggregate.

Figure 4:      Fit of the aggregate form factor with the two-level fractal model ($G_1$ = 9550 cm$^{-1}$, $R_g$ = 1650 Å, d = 1.96). The one-level model is not a fit. Its parameters ($G_2$ = 103 cm$^{-1}$, radius of gyration R = 76 Å, Porod decay p = -4) have been taken from the form factor of individual silica beads.

Figure 5:      Evolution of $\chi^2$ with the number of Monte Carlo tries per bead for three different initial conditions. In the inset, a long run with 300 tries per bead.

Figure 6:      Graphical representations of aggregate structures. Two initial configurations are shown on the left. The structures on the right are snapshots after 300 (top) and 30 (bottom) tries per bead, each starting from the initial configurations on the left.

Figure 7:      Structure of silica-latex nanocomposite ($\circ$, $\Phi = 5\%$, pH 7, B30) compared to the RMC model prediction ($N_{agg} = 93$, solid line). In the lower inset the aggregate form factor is compared to the RMC result. In the upper inset, the RMC-results ($P_{agg}$) for higher aggregation numbers ($N_{agg} = 120$ and 150, solid lines) are compared to the nominal one ($N_{agg} = 93$, symbols).

Figure 8:      SANS-intensities of samples (B30 pH5) with silica volume fraction of 5% and 10% (symbols). The solid lines are the RMC-results. For illustration, the intensity of the RMC-algorithm calculated from the initial aggregate configuration is also shown (10%). In the inset, aggregate form factors $P_{agg}$ are compared.

Figure 9:      Structure of silica-latex nanocomposites (symbols, $\Phi = 3\%$-15, pH 5, B40) compared to the RMC model predictions (see text for details).

Figure 10:    Snapshots of aggregate structures at different silica volume fractions as calculated by RMC (pH5, B40-series).

**Tables**

| Φ | Debye length factor | Charge | $N_{agg}$ |
|---|---|---|---|
| 5% | 60% | 61 | 430 |
| 10% | 175% | 52 | 309 |

Table 1: Parameters used for a successful decomposition in $S_{inter}$ and an artefact-free $P_{agg}$, for series B30, pH5. The Debye length is given as a multiple of the surface-to-surface distance between neighboring aggregates.

| Φ | Debye length factor | Charge | $N_{agg}$ |
|---|---|---|---|
| 3% | 120% | 52 | 120-200 |
| 6% | 150% | 58 | 168 |
| 9% | 150% | 78 | 196 |
| 12% | 250% | 63 | 238 |
| 15% | 275% | 55 | 292 |

Table 2: Parameters used for a successful decomposition in $S_{inter}$ and an artefact-free $P_{agg}$, for series B40, pH5. The Debye length is given as a multiple of the surface-to-surface distance between neighboring aggregates.

| Φ | d | $R_g$ (Å) fractal | $R_g$ (Å) RMC | D (Å) | $E/E_{latex}$ |
|---|---|---|---|---|---|
| 3% | 1.6 | 3470 | 2830 | 2400 | 2.8 |
| 6% | 2.0 | 2640 | 1690 | 2000 | 6.4 |
| 9% | 2.2 | 2290 | 2090 | 1780 | 23.2 |
| 12% | 2.3 | 2150 | 1870 | 1750 | 29.6 |
| 15% | 2.4 | 2550 | 2680 | 1750 | 42.5 |

Table 3: Series B40, pH5. Fractal dimension d and radius of gyration $R_g$ from one-level Beaucage fit compared to $R_g$ determined by RMC and inter-aggregate distance from $S_{inter}$. The last column recalls the mechanical reinforcement factor of these samples.

**Figures**

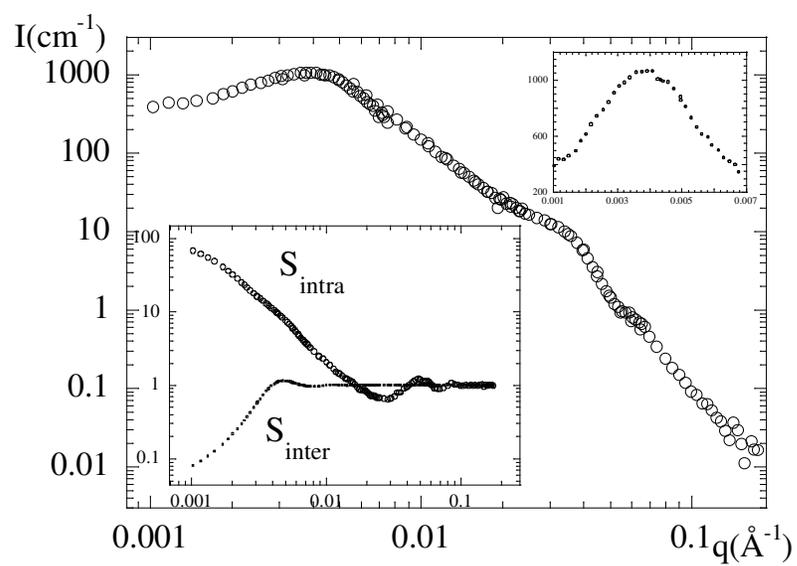

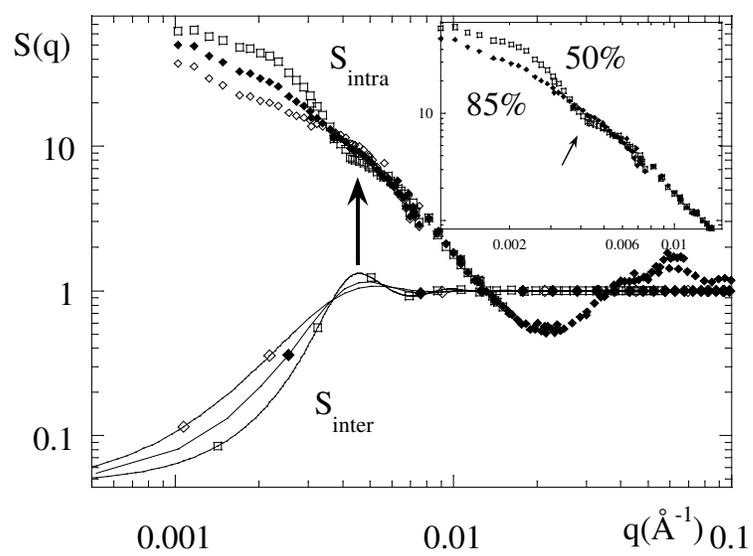

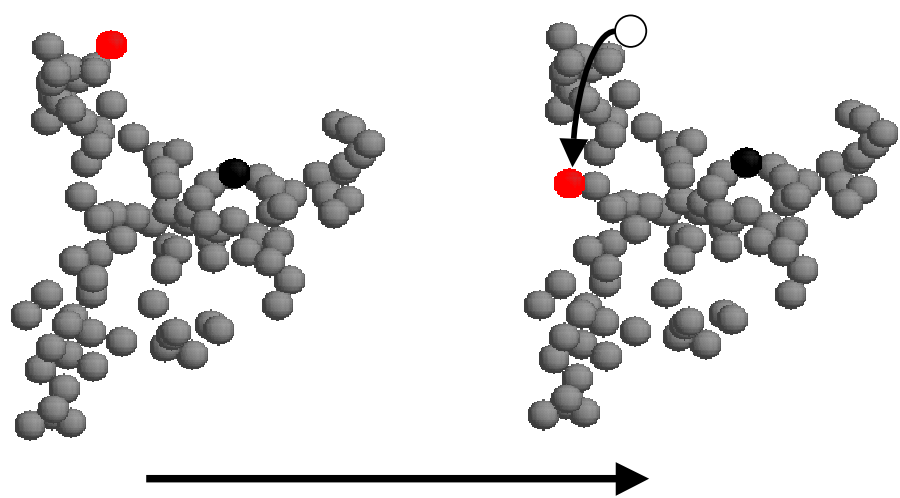

F<small>IGURE</small> 3 (O<small>BERDISSE</small>)

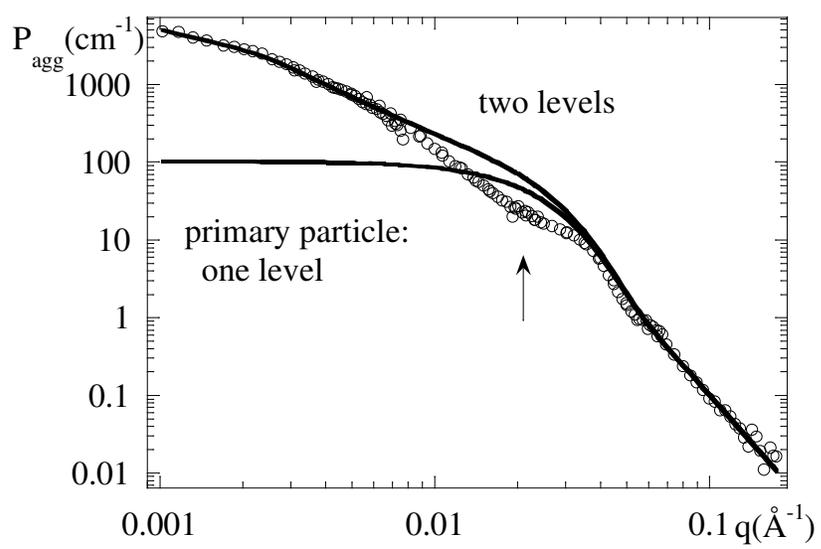

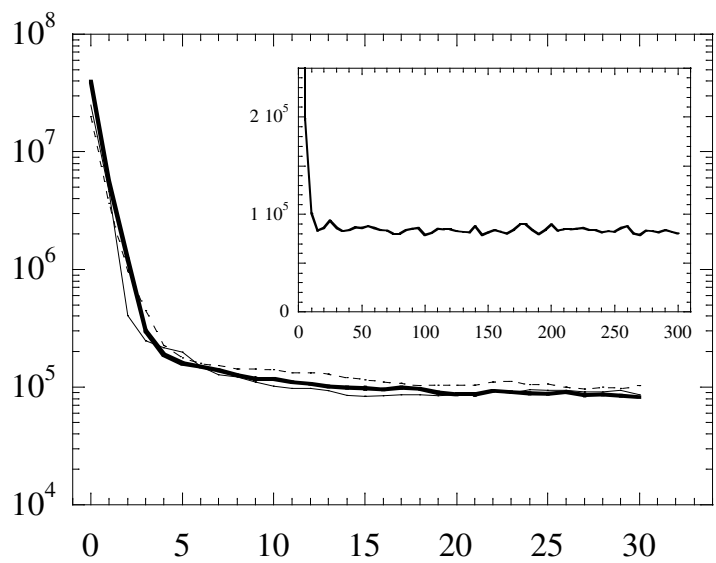

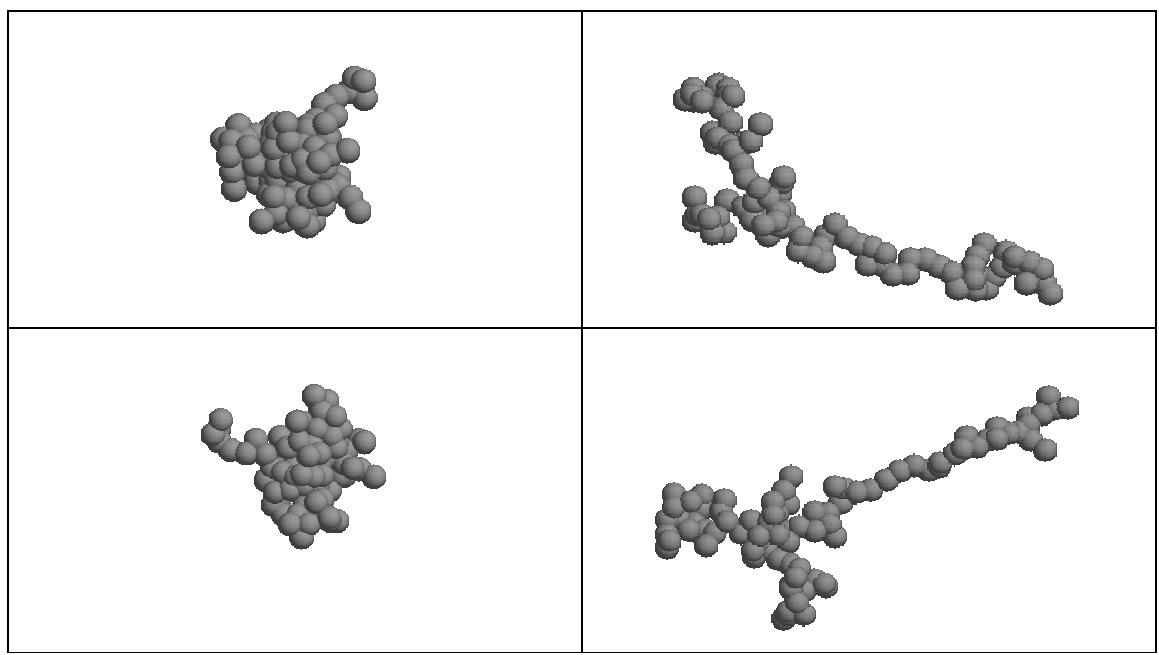

FIGURE 6 (OBERDISSE)

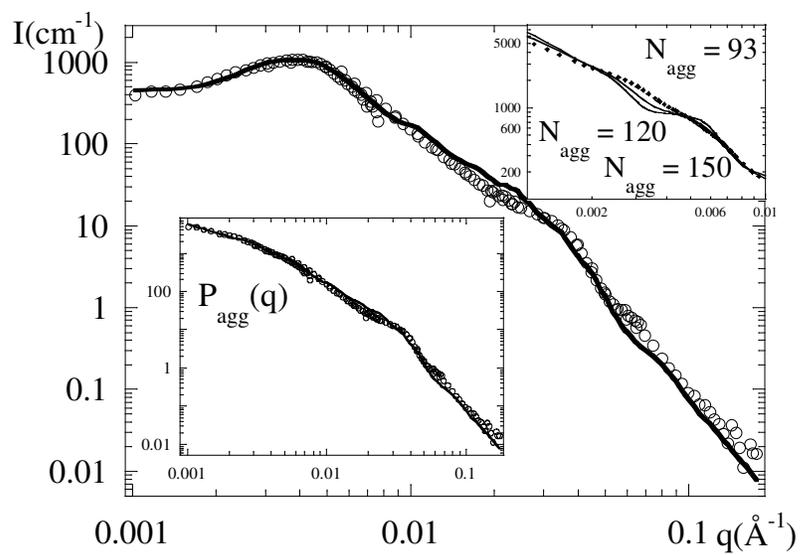

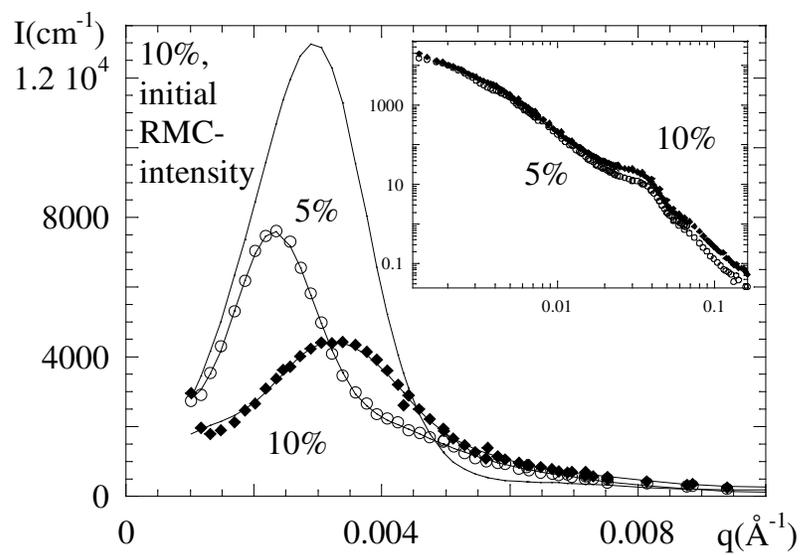

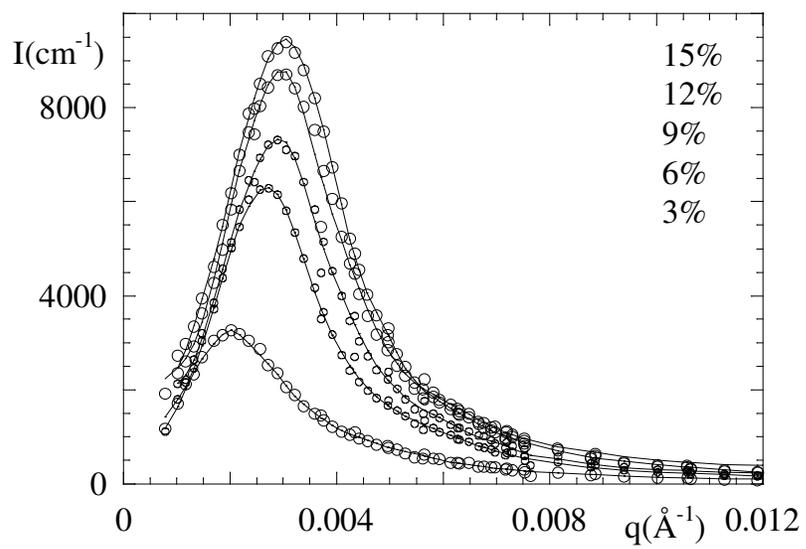

FIGURE 9 (OBERDISSE)

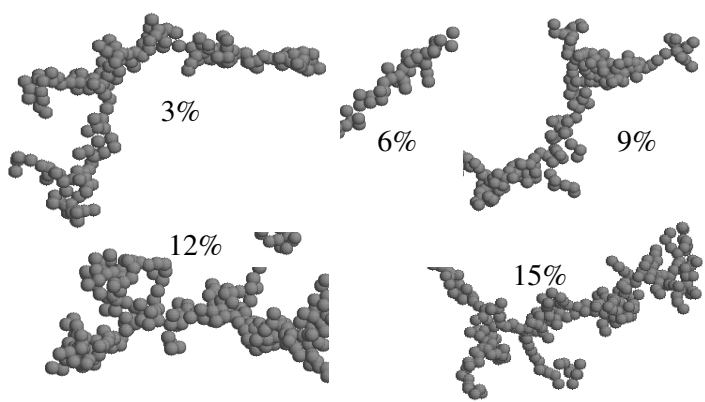